# Performance Analysis of a Near-Field Thermophotovoltaic Device with a Metallodielectric Selective Emitter and Electrical Contacts for the Photovoltaic Cell

Yue Yang, Jui-Yung Chang, and Liping Wang [a]

School for Engineering of Matter, Transport, and Energy
Arizona State University, Tempe, AZ, USA 85287

**ABSTRACT**

A near-field thermophotovoltaic (TPV) system with a multilayer emitter of alternate tungsten and alumina layer is proposed in this paper. The fluctuational electrodynamics along with the dyadic Green's function for a multilayered structure is applied to calculate the spectral heat flux, and the charge transport equations are solved to get the photocurrent generation and electrical power output. The spectral heat flux is much enhanced when plain tungsten emitter is replaced with multilayer emitter. The mechanism of surface plasmon polariton coupling in the tungsten thin film, which is responsible for the heat flux enhancement, is analyzed. In addition, the invalidity of effective medium theory to predict the optical properties of multilayer structure in near-field radiation is discussed. The tungsten and alumina layer thicknesses are optimized to match the spectral heat flux with the bandgap of TPV cell. Practically, with a gold reflector placed on the back of TPV cell, which also acts as the back electrode, and a 5-nm-thick indium tin oxide (ITO) layer as the front contact, when the emitter and receiver temperature are respectively set as 2000 K and 300 K, the conversion efficiency and electrical power output can be achieved to 23.7% and 0.31 MW/m$^2$ at a vacuum gap distance of 100 nm.

[a] Corresponding author. Contact information: liping.wang@asu.edu

# 1. INTRODUCTION

TPV energy conversion system, which consists of two main components of thermal emitter and TPV cell, is a direct energy conversion process from heat to electricity through electron-hole pair generation in TPV cell from radiation absorbing. The advantages of TPV system such as versatile thermal sources, high power output and conversion efficiency potential [1], and few moving parts, make it possible to have wide applications in space, military area and microelectronics. However, the biggest challenge for TPV system right now is the low power output and conversion efficiency.

Recently, it has been demonstrated that the heat transfer in near field can be greatly enhanced and exceed the blackbody limit for few orders due to the coupling of evanescent waves and surface waves [2-6]. In order to improve the power output, the idea of applying near-field radiation enhancement to TPV system was proposed by Whale [7], in which blackbody emitter and extra recombination loss were considered. Laroche also developed a model of near-field TPV considering a tungsten emitter with TPV cell of 100% quantum efficiency [8]. Basu presented a review of the near-field TPV system in 2007 [1]. The same group also constructed a near-field TPV system using plain tungsten as the thermal emitter, and they considered for the first time the practical energy absorption and charge transport in different regions of the TPV cell [9]. In order to enhance the conversion efficiency, a gold reflector was added at the back of TPV cell to reflect the useless long wavelength energy back to emitter [10]. More recently, Francoeur investigated the thermal impacts on the performance of near-field TPV system by solving coupled near-field thermal radiation, charge transport, and heat transport formulations in TPV cell, and concluded that the TPV cell temperature will affect the performance greatly [11].

Efforts have been made to design selective emitters for far-field TPV systems to improve the conversion efficiency. Due to the narrow band radiation, rare-earth oxides were the earliest selective emitters [12, 13]. Benefiting from the improvement of micro/nanoscale fabrication technique, micro/nanostructured surfaces were also used to achieve selective emission. The grating structures with different dimensions were considered as the most widely applicable selective emitters and filters [14-16]. Magnetic polaritons, which account for the strong coupling of external electromagnetic waves with the magnetic resonance excited inside the nanostructure, is another mechanism used to achieve selective emission [17-19]. Recently, the metamaterials using epsilon-near-zero (ENZ) or epsilon-near-pole (ENP) have also been proposed as selective emitter [20]. Narayanaswamy showed that a stack of alternate tungsten layer and alumina layer could be applied as the emitter of far-field TPV system in 2004 [21]. One question is proposed that whether the emission spectrum can be spectrally tuned to enhance the conversion efficiency for near-field TPV system. Narayanaswamy [22] presented the idea to improve the performance of near-field TPV by exciting SPhP coupling between dielectric emitter and optimized TPV cell. Several groups have also recently used the property of graphene sheet to tune the surface plasmon resonance frequency of either the emitter or TPV cell to make both the surface plasmon frequencies to couple with each other [23, 24]. The thin-film germanium emitter was also proposed as thermal well to achieve spectrally enhanced near-field radiative heat flux [25]. The Drude radiator, which supports surface polaritons in near infrared region and aims to achieve quasi-monochromatic radiative heat flux, was also applied as the emitter of near field TPV system [26], and the impacts of radiative, electrical and thermal losses on near-field TPV performance were investigated as well [27]. The tungsten nanowire based hyperbolic metamaterial was also applied as near-field TPV emitter to improve the performance due to the supported hyperbolic modes [28].

In this study, a multilayer emitter with alternate tungsten layer and alumina layer, which was proposed by Narayanaswamy [21] for far-field TPV system, will be studied for a near-field TPV system with a TPV cell made of $In_{0.18}Ga_{0.82}Sb$ [9]. The optical properties of both tungsten and alumina are obtained from Palik's data [29]. $In_{0.18}Ga_{0.82}Sb$ is an alloy of InSb and GaSb, and the temperature dependent optical properties at different volume fractions are considered [29, 30]. The bandgap of this alloy is $E_g$ = 0.56 eV, which corresponds to $\lambda_g$ = 2.22 μm. As depicted in Fig. 1, tungsten is set as the first layer of emitter. The temperatures of the emitter and the TPV cell are set as 2000 K and 300 K, respectively. The near-field heat flux between the multilayer emitter and the TPV cell is calculated, and the mechanisms on the radiative transport will be investigated. The TPV system power output and conversion efficiency is also analyzed through solving charge transport equations. In order to reflect the long wavelength waves with energy smaller than the bandgap of TPV cell back to emitter to enhance the conversion efficiency, a gold reflector is placed on the back of TPV cell. Note that the gold reflector can also act as the back electrode to extract the electricity. Moreover, an ITO thin film, which serves as the front transparent electrode, is added at the front of TPV cell.

## 2. THEORETICAL BACKGROUND

### 2.1 Near-Field Radiative (NFR) Transport between Multilayer Isotropic Structures

The fluctuational electrodynamics theory along with the dyadic Green's function for a multilayered structure is applied to calculate the radiative heat flux between multilayers. The electrical field **E** and magnetic field **H** can be expressed as follows [31].

$$\mathbf{E}(\mathbf{x},\omega) = i\omega\mu_0 \int_V \bar{\bar{\mathbf{G}}}_e(\mathbf{x},\mathbf{x}',\omega) \cdot \mathbf{j}(\mathbf{x}',\omega) d\mathbf{x}' \tag{1a}$$

$$\mathbf{H}(\mathbf{x},\omega) = \int_V \nabla \times \bar{\bar{\mathbf{G}}}_e(\mathbf{x},\mathbf{x}',\omega) \cdot \mathbf{j}(\mathbf{x}',\omega) d\mathbf{x}' \tag{1b}$$

where $\mu_0$ is the magnetic permeability in vacuum, and $i$ is unit of purely imaginary number. $\bar{\bar{\mathbf{G}}}_e(\mathbf{x},\mathbf{x}',\omega)$ is the electric dyadic Green's function, which presents the relationship between the space- and time- dependent electric current density $\mathbf{j}(\mathbf{x}',\omega)$ at location $\mathbf{x}'$ and at the angular frequency $\omega$, which is due to charge movement from thermal fluctuation, and the electric field $\mathbf{E}(\mathbf{x},\omega)$ at location $\mathbf{x}$. The expressions of electric dyadic Green's function for a multilayered structure have been well derived [9, 21, 32-36]. The spectral heat flux at location $\mathbf{x}$ can be expressed as the time-averaged Poynting vector.

$$\mathbf{S}_\omega(\mathbf{x},\omega) = \frac{1}{2}\left\langle \text{Re}[\mathbf{E}(\mathbf{x},\omega) \times \mathbf{H}^*(\mathbf{x},\omega)] \right\rangle \tag{2}$$

where the superscript * denotes the conjugate of complex number.

Therefore, the total heat flux reaching the surface at the location of $z_m$ in the z direction can be obtained by integrating the spectral heat flux.

$$q''(z_l) = \int_0^\infty S_{z,\omega}(z_m,\omega) d\omega \tag{3}$$

Then the energy absorbed by a layer is calculated as the difference between the total heat fluxes reaching the two surfaces of this layer.

**2.2 Charge Transport in the TPV Cells**

Here the photocurrent generation is considered separately in different regions of a TPV cell with charge transport equations. The TPV cell is separated into N layers, and with the spectral heat flux absorbed by $l$th layer with thickness of $d_l$, denoted as $Q_{\lambda,l}$, the electron-hole pairs generation rate in this layer is described as [9]

$$g_l(\lambda) = \frac{Q_{\lambda,l}}{d_l(hc_0/\lambda)} \text{ with } \frac{hc_0}{\lambda} \geq E_g \tag{4}$$

where $h$ is the Planck constant, $c_0$ is the light velocity in vacuum and $hc_0/\lambda$ is the photon energy at wavelength $\lambda$. $E_g$ is the bandgap of TPV cell. This equation means that only the photons with energy higher than $E_g$ can generate electron-hole pairs. $d_l$ is the thickness of the $l$th layer in the receiver.

In depletion region of the TPV cell, the generated electron-hole pairs can be completely collected due to the built-in voltage, and the drift current is expressed as

$$J_{dp}(\lambda) = e g_{dp}(\lambda) L_{dp} \tag{5}$$

where $e$ is the electron charge, and $L_{dp}$ is the depletion region thickness.

The drift currents generated in p-region and n-region, which are denoted as $J_p(\lambda)$ and $J_n(\lambda)$, can be attained by solving the charge transport equations with the boundary conditions. The total generated drift current $J_\lambda(\lambda)$ is the summation of the three drift currents $J_\lambda(\lambda) = J_{dp}(\lambda) + J_p(\lambda) + J_n(\lambda)$. The quantum efficiency $\eta_q(\lambda)$, which is the ratio of the number of generated electron-hole pairs and the number of absorbed photons, can be obtained through the total generated drift current and absorbed spectral heat flux [9].

$$\eta_q(\lambda) = \frac{J_\lambda / e}{S_\lambda / (hc/\lambda)} \tag{6}$$

where $S_\lambda$ is the spectral heat flux absorbed by the whole TPV receiver.

The electrical power output can be expressed as [9]

$$P_E = J_{ph} V_{oc} (1 - 1/y)[1 - \ln(y)/y] \tag{7}$$

where $y = \ln(J_{ph}/J_o)$. $J_{ph}$, $V_{oc}$, $J_o$ are photocurrent, short circuit voltage, and dark current, respectively. The expressions of them have been described in Ref. [15]. The important parameter to characterize the performance of TPV system, the conversion efficiency $\eta$, is expressed as the ratio of electrical output and radiative power input.

$$\eta = P_E / P_R \qquad (8)$$

where $P_R = \int_0^\lambda S_\lambda d\lambda$ is the total absorbed radiative power by the receiver.

For the TPV cell, the same parameters were used with Ref. [15]. The thicknesses of p-region, depletion region, and n-region are set as 0.4 μm, 0.1 μm, and 10 μm, respectively. The p-region has a doping concentration of $10^{19}$ cm$^{-3}$, and that for n-region is $10^{17}$ cm$^{-3}$. The diffusion coefficient and relaxation time for electron in p-region are set as 125 cm$^2$s$^{-1}$ and 9.75 ns, respectively, while those for hole in n-region are set as 31.3 cm$^2$s$^{-1}$ and 30.8 ns, respectively. The surface recombination velocities on the top surface of p-region and the bottom surface of n-region are set as $7.4\times10^4$ ms$^{-1}$ and 0 ms$^{-1}$, respectively. Recombination on the bottom surface of n-region is not considered because the thickness of n-region is much larger than hole diffusion length.

## 3. RESULTS AND DISCUSSIONS

### 3.1 Spectral Heat Flux between the Multilayer Emitter and the TPV Cell

The multilayer emitter is semi-infinite with alternate tungsten and alumina layer, which is shown in Fig. 1. A total 20 layers structure is used for simulation, of which the convergence has been checked by limiting the relative difference of total heat flux smaller than 0.8% compared to that of 60 layers structure.

The spectral heat fluxes between the multilayer emitter and the TPV cell for different tungsten and alumina layer thickness are presented in Fig. 2. The vacuum gap distance is $d = 100$ nm. For Fig. 2(a), the tungsten layer thickness is fixed, and the alumina layer thickness varies. The spectral heat flux between plain tungsten emitter and the TPV cell indicated by the black dash line is also plotted for comparison. It can be clearly seen that the spectral heat flux is much enhanced when the plain tungsten emitter is replaced by multilayer emitter. As a good selective

TPV emitter, it should emit more energy above the bandgap $E_g$, which means the wavelength smaller than $\lambda_g = 2.22$ μm, and less energy below the bandgap to maximize both the power output and conversion efficiency. Therefore, comparing the results of spectral heat flux for different alumina layer thickness, the optimum thickness is chosen as $t_{Al2O3} = 300$ nm. With the optimum thickness ratio of tungsten and alumina layer as $t_W/t_{Al2O3} = 1/30$ obtained from Fig. 2(a), the spectral heat fluxes for different tungsten layer thicknesses are plotted in Fig. 2(b). Based on the same principle to achieve the best TPV performance, the multilayer emitter structure with tungsten layer thickness of $t_W = 10$ nm and alumina layer thickness of $t_{Al2O3} = 300$ nm is applied via comparing with other thicknesses combinations discussed above. All the results shown about the multilayer emitter in the following paragraphs are based on these geometric dimensions.

### 3.2 Comparison with Effective Medium Theory (EMT)

Using the Maxwell-Garnett EMT[37, 38], the multilayer emitter can be approximated as a homogeneous and uniaxial medium with component of dielectric function parallel to the surface $\varepsilon_\parallel$ and vertical component $\varepsilon_\perp$ expressed as follows.

$$\varepsilon_\parallel = f\varepsilon_m + (1-f)\varepsilon_d \tag{9a}$$

$$\varepsilon_\perp = \frac{\varepsilon_m \varepsilon_d}{f\varepsilon_d + (1-f)\varepsilon_m} \tag{9b}$$

where $f$ is the volume fraction of metal in the whole structure, and $\varepsilon_m$ and $\varepsilon_d$ are the dielectric functions of metal and dielectric medium, respectively. In order to understand the mechanism to result in the heat transfer enhancement, the spectral heat fluxes are separated into different polarizations, and the comparison between the spectral heat fluxes calculated from multilayer NFR and EMT is investigated in Fig. 3. Here, fluctuational electrodynamics [31] along with wave propagation in anisotropic medium, is used to calculate the near-field radiative heat fluxes between the effective uniaxial medium emitter with the dielectric function defined in Eq. (9), and the receiver, of which the details can be found in Ref. [39].

As can be seen from Fig. 3, there is a large deviation between the spectral heat fluxes obtained using EMT and multilayer NFR for TE waves. As we all know, the basic condition to use EMT is that the characteristic wavelength should be at least 10 times larger than the structure size. However, the characteristic wavelength, which is around 1.5 μm from Fig. 2, is only about 5 times of the period of multilayer, which is 310 nm. The small ratio between characteristic wavelength and the multilayer structure period results in the invalidity of EMT to predict heat transfer for TE waves. Furthermore, when it comes to TM waves, the most significant deviation between the spectral heat fluxes obtained from exact multilayer NFR and EMT is around $\lambda_{SPP}$ = 2 μm, where there is a high peak of heat flux from multilayer NFR calculation, while the peak cannot be observed from EMT prediction. The spectral heat flux peak at around $\lambda_{SPP}$ = 2 μm for TM waves is caused by the surface plasmon polaritons (SPP) coupling within the tungsten thin film, which will be verified later. This also proves that EMT cannot predict SPP coupling among multilayer structure at small vacuum gap. Actually, the invalidity of EMT for multilayer structures in the presence of surface waves has been discussed in details previously [40]. It even showed that different topmost layer can result in different density of states above the multilayer structure, which cannot be captured by EMT. The effects of finite number of multilayer stacks, defects and surface roughness during practical fabrication process on the accuracy of EMT prediction were investigated as well [41]. Moreover, Zhang's group quantitatively presented the application condition of EMT to predict the near-field radiative heat transfer between multilayer structure recently [42].

## 3.3 Elucidating the Mechanisms for Spectral Near-field Radiation Enhancement at TM Polarization

As mentioned above, the spectral heat flux increase for the TM waves of multilayer emitter, which is around $\lambda_{SPP} = 2$ μm shown in Fig. 3, is due to the SPP coupling within the tungsten thin film. In order to verify this hypothesis, the contour plot of energy transmission coefficient for TM waves ($\xi$) between the multilayer emitter and receiver is made in Fig. 4 for multilayer NFR calculation. The energy transmission coefficient, which only considers the material properties of emitter and receiver, is the function of both angular frequency $\omega$ and parallel wavevector $\beta$ and the definition can be found in Ref. [43]. The relationship between spectral heat flux ($S_\omega$) and transmission coefficient ($\xi$) can be expressed as

$$S_\omega = \frac{1}{4\pi^2}\left[\theta(T_H,\omega) - \theta(T_L,\omega)\right]\int_0^\infty \beta\xi(\omega,\beta)d\beta \tag{10}$$

where $\theta(w,T) = \hbar w/[\exp(\hbar w/k_B T) - 1]$ is the mean energy of a Planck oscillator at thermal equilibrium temperature $T$, where $\hbar$ and $k_B$ are the Planck constant divided by $2\pi$ and the Boltzmann constant, respectively.

The extra energy transmission coefficient enhancement can be clearly observed at small angular frequencies. Note that the energy transmission coefficient enhancement at $\omega = 9\times10^{14}$ rad/s in Fig. 4 exactly corresponds to the heat flux peak at $\lambda_{SPP} = 2$ μm shown in Fig. 3. The coupled SPP dispersion curve for TM waves within the tungsten thin film with vacuum substrate on one side and alumina substrate on the other side is also plotted in Fig. 4. The equation to plot the dispersion curve is as follows [44].

$$\tanh(ik_{2z}t_2)\left(\frac{k_{2z}^2}{\varepsilon_2^2} + \frac{k_{1z}k_{3z}}{\varepsilon_1\varepsilon_3}\right) = \frac{k_{2z}}{\varepsilon_2}\left(\frac{k_{1z}}{\varepsilon_1} + \frac{k_{3z}}{\varepsilon_3}\right) \tag{11}$$

where the subscripts 1, 2, 3 denote the vacuum substrate, tungsten thin film, and alumina substrate, respectively. $k_{jz} = \sqrt{\varepsilon_j \omega^2 / c_0^2 - \beta^2}$ ($j$ = 1, 2, 3) is the component of wavevector vertical to the interface, and $\beta$ is the wavevector parallel to the interface. $t_2$ is the thickness of tungsten thin film. The perfect match between the coupled SPP dispersion curve and the energy transmission coefficient enhancement for TM waves in Fig. 4 further verifies the hypothesis that the spectral heat flux enhancement when applying multilayer emitter results from the SPP coupling within the tungsten thin film.

In order to understand the contribution of emission from each tungsten layer, the spectral heat flux with contributions of both TE and TM waves from different layer of the multilayer emitter is plotted in Fig. 5. When considering the emission from a single layer, the others just act as filters. The 1st layer is the bottom tungsten layer of the emitter as depicted in Fig. 1. From Fig. 5, it can be observed that the spectral heat flux from the bottom tungsten layer is much higher than the other layers, even compared to the tungsten layer just above it, which is because waves decay very fast in the multilayer emitter, especially for exponentially decayed evanescent waves. The spectral heat flux peak around $\lambda_{ba}$ = 1.5 μm at each tungsten layer is due to the band absorption of tungsten, which is contributed by both TE and TM waves, while the peak around $\lambda_{SPP}$ = 2 μm is due to SPP coupling within tungsten thin film only for TM waves. Therefore, as a conclusion, for multilayer emitter with alternate tungsten and alumina layer, the bottom tungsten layer contributes the most energy emission, and the SPP coupling in each tungsten thin film greatly enhances the heat flux.

**3.4 Performance of the Near-field TPV System at Different Vacuum Gap Distances**

Calculated from the Eq. (6), the quantum efficiencies against wavelength at different vacuum gap distances are shown in Fig. 6(a). When the wavelength is larger than $\lambda_g$ = 2.22 μm, which means the photon energy is below bandgap, no electron-hole pair is generated, so the quantum efficiency is zero. On the other hand, for smaller wavelength, most of the photons will be absorbed close to the surface because of its small penetration depth. Thus, due to the existence of strong recombination at the surface, the quantum efficiency also decreases for small wavelength. The quantum efficiency is around $\eta_q = 80\%$ within the wavelength band from 1 μm to 2 μm. Note that, it has smaller quantum efficiency with a 10 nm vacuum gap because of the small penetration depth.

Integrating the spectral photocurrent density over the wavelength, the photocurrent density generated in different regions of TPV cell as a function of vacuum gap distance is obtained in Fig. 6(b). From this figure, most of the photocurrent is generated in n-region because of the large thickness and that most of the energy is absorbed in n-region. When the vacuum gap distance decreases, more percentages of photocurrent is generated in p-region, because smaller gap distance means smaller penetration depth and more energy will be absorbed in p-region.

The electrical power output and radiative power input of the TPV system, indicated in Eq. (7) and Eq. (8), for both multilayer emitter and plain tungsten emitter are plotted as a function of vacuum gap distance in Fig. 7(a). Because of the enhanced spectral heat flux using multilayer emitter indicated in Fig. 2, the radiative power input is enhanced compared to the plain tungsten emitter. Moreover, the electrical power output is improved as well due to the enhanced spectral heat flux above bandgap. As the ratio of electrical power output and radiative power input, the conversion efficiency is plotted in Fig. 7(b). As the increase rate of radiative power input is larger than that of electrical power output, the conversion efficiency of the TPV system with a multilayer emitter is lower than that with a plain tungsten emitter. Therefore, a higher electrical power output of about $P_E = 0.34$ MW/m$^2$ and a lower conversion efficiency of $\eta$

= 17.2% of TPV system with multilayer emitter at 100 nm gap are achieved compared to that with plain tungsten emitter.

**3.5 Performance Improvement by Adding Au Back Reflector and ITO Front Contact for the TPV Cell**

As Fig. 7 shows, compared with plain tungsten emitter, the conversion efficiency is decreased using multilayer emitter, which is due to higher energy waste on radiative heat flux below the bandgap as shown in Fig. 3. In order to decrease this fraction of heat flux to increase the conversion efficiency, a gold reflector on the back of n-region layer is applied to reflect the long wavelength heat flux back to the emitter. The optical properties of gold is obtained from Palik's tabular data [29]. In reality, the gold reflector also works as the back electrode to extract the current. Considering both the aspects of good electrical conductance and photon transparency in near-infrared region, the transparent conducting oxide material ITO is chosen as the front electrode. Here the thickness of ITO layer is set as 5 nm, which is still available using nano-fabrication technique and also minimizes the absorption in ITO layer. The dielectric function of ITO in near-infrared region can simply be described by the Drude model [45].

Figure 8 shows different results for three different receivers: semi-infinite TPV cell, TPV cell with opaque Au beneath it, and TPV cell with both opaque Au beneath it and 5-nm-thick ITO on top of it, respectively. Note that multilayer emitter is assumed for all the three cases. The spectral heat fluxes absorbed by the whole receiver at the vacuum gap distance of 100 nm for the three cases are shown in Fig. 8(a). Comparing all the three results, the energy absorbed above the bandgap is almost the same, but the energy absorbed below the bandgap is greatly decreased with the gold reflector. This indicates that lower radiative power input can be achieved with same electrical power output. On the other hand, as can be seen from Fig. 8(a), adding a 5-nm-thick ITO layer on top of TPV cell has minimal effect on spectral heat flux with 100 nm vacuum gap. However, with a vacuum gap distance of 10 nm as shown in Fig. 8(b), a very sharp and high heat flux peak at the wavelength around $\lambda = 1.2$ μm is observed when adding an ITO front layer. This is because ITO is highly absorbing material at the wavelength beyond 1 μm, and most of the energy with small penetration depth due to small vacuum gap distance and small wavelength will be absorbed by ITO thin film.

The radiative power input and electrical power output for all the three cases are plotted in Fig. 9(a). As expected from Fig. 8(a) and Fig. 8(b), a smaller radiative power input and almost the same electrical power output can be achieved when adding a gold reflector. However, different from gold reflector, adding ITO front layer will greatly increase the radiative power input at small vacuum gap distances. And because at small vacuum gap distances, the enhanced energy input is mainly absorbed in the ITO thin film, the electrical power output remains almost the same.

Calculating the ratio of electrical power output and radiative power input, the conversion efficiency is plotted in Fig. 9(b). Because of the decreased radiative power input, much higher conversion efficiency is achieved when using gold reflector. At the vacuum gap of $d = 100$ nm, the conversion efficiency of $\eta = 16.6\%$ for the case of multilayer emitter without gold reflector is increased to $\eta = 25.6\%$ when using gold reflector, which is also higher than that for plain tungsten emitter case ($\eta = 20.0\%$). Therefore, compared to the plain tungsten emitter, both the electrical power output and conversion efficiency are improved when using multilayer emitter with gold reflector. In addition, more practically, when adding a 5-nm-thick ITO layer as the front electrode, although conversion efficiency will decrease because of the higher radiative power input at vacuum gap distance smaller than 50 nm, a high conversion efficiency of 23.7% is still achieved at $d = 100$ nm.

## 4. CONCLUSION

A multilayer emitter with alternate tungsten and alumina layer and the TPV cell of $In_{0.18}Ga_{0.82}Sb$ are applied to construct a near-field TPV system. The spectral heat flux is much enhanced by replacing the plain tungsten emitter with multilayer emitter due to the SPP coupling within tungsten thin film, and the thicknesses of tungsten and alumina layer are respectively optimized as $t_W = 10$ nm and $t_{Al2O3} = 300$ nm to match the bandgap of TPV cell. And the invalidity of EMT to predict the property of structure with large period and the nonlocal effects such as SPP coupling has been verified. By considering the charge transport in different regions of TPV cell, the photocurrent density in different regions and the electrical power output are calculated. With an opaque gold reflector beneath the TPV cell and a 5-nm-thick ITO layer as the front contact, the conversion efficiency of $\eta = 23.7\%$ and electrical power output of $P_E = 0.31$ MW/m$^2$ can be achieved at the vacuum gap of $d = 100$ nm when the emitter and receiver temperatures are set as $T_H = 2000$ K and $T_L = 300$ K, respectively. The experimental work of this near-field TPV system and searching for another front contact instead of ITO whose drawback is the highly absorbing property at larger wavelength will be explored in the future.

**ACKNOWLEDGMENT**

The authors are grateful to the supports from ASU New Faculty Startup fund. YY would like to thank the partial support from the University Graduate Fellowship offered by the ASU Fulton Schools of Engineering.

# NOMENCLATURE

| | |
|---|---|
| $c_0$ | light velocity in vacuum, $2.997 \times 10^8$ m s$^{-1}$ |
| $d_l$ | thickness of $l$th layer in receiver (m) |
| **E** | electric field vector (V/m) |
| $E_g$ | bandgap of TPV cell (eV) |
| $e$ | charge of electron, $1.6 \times 10^{-19}$ C |
| $f$ | filling ratio of metal in multilayer structure |
| $\bar{\bar{\mathbf{G}}}_e$ | electric-field dyadic Green's function (m$^{-1}$) |
| $g$ | electron-hole generation rate |
| **H** | magnetic field vector (A/m) |
| $h$ | Planck constant, $6.626 \times 10^{-34}$ m$^2$ kg s$^{-1}$ |
| $J$ | drift current density (A m$^{-2}$) |
| $J_{ph}$ | short circuit current (A m$^{-2}$) |
| **j** | fluctuating current density (A m$^{-2}$) |
| $k_z$ | wavevector vertical to interface |
| $L$ | thickness of different region in TPV cell (m) |
| $P_E$ | electrical power output (W m$^{-2}$) |
| $P_R$ | radiative power input (W m$^{-2}$) |
| $Q_{\lambda,l}$ | spectral heat flux absorbed in $l$th layer (W m$^{-2}$ μm$^{-1}$) |
| $q''$ | radiative heat flux (W m$^{-2}$) |
| $S_\omega$ | spectral Poynting vector (W s m$^{-2}$ rad$^{-1}$)) |
| $S_{z,\omega}$ | spectral heat flux in z direction (W s m$^{-2}$ rad$^{-1}$) |
| $T$ | temperature (K) |
| $V_{oc}$ | open circuit voltage (V) |
| **x** | space variable (m) |
| **x'** | space variable (m) |
| $z_m$ | z position of $m$th layer (m) |

**GREEK**

| | |
|---|---|
| $\beta$ | wavevector parallel to the interface |
| $\varepsilon$ | relative electric permittivity |
| $\varepsilon_d$ | relative electric permittivity of dielectric |
| $\varepsilon_m$ | relative electric permittivity of metal |
| $\omega$ | angular frequency (rad s$^{-1}$) |
| $\lambda$ | wavelength (m) |
| $\mu_0$ | magnetic permeability in vacuum, $4\pi \times 10^{-7}$ N A$^{-2}$ |
| $\eta$ | conversion efficiency |
| $\eta_q$ | quantum efficiency |
| $\xi$ | energy transmission coefficient |

**SUPERSCRIPTS**

| | |
|---|---|
| $*$ | complex conjugate |

**SUBSCRIPTS**

| | |
|---|---|
| $\perp$ | direction vertical to the surface |
| ‖   | direction parallel to the surface |
| dp  | depletion region of TPV cell |
| n   | n region of TPV cell         |
| p   | p region of TPV cell         |

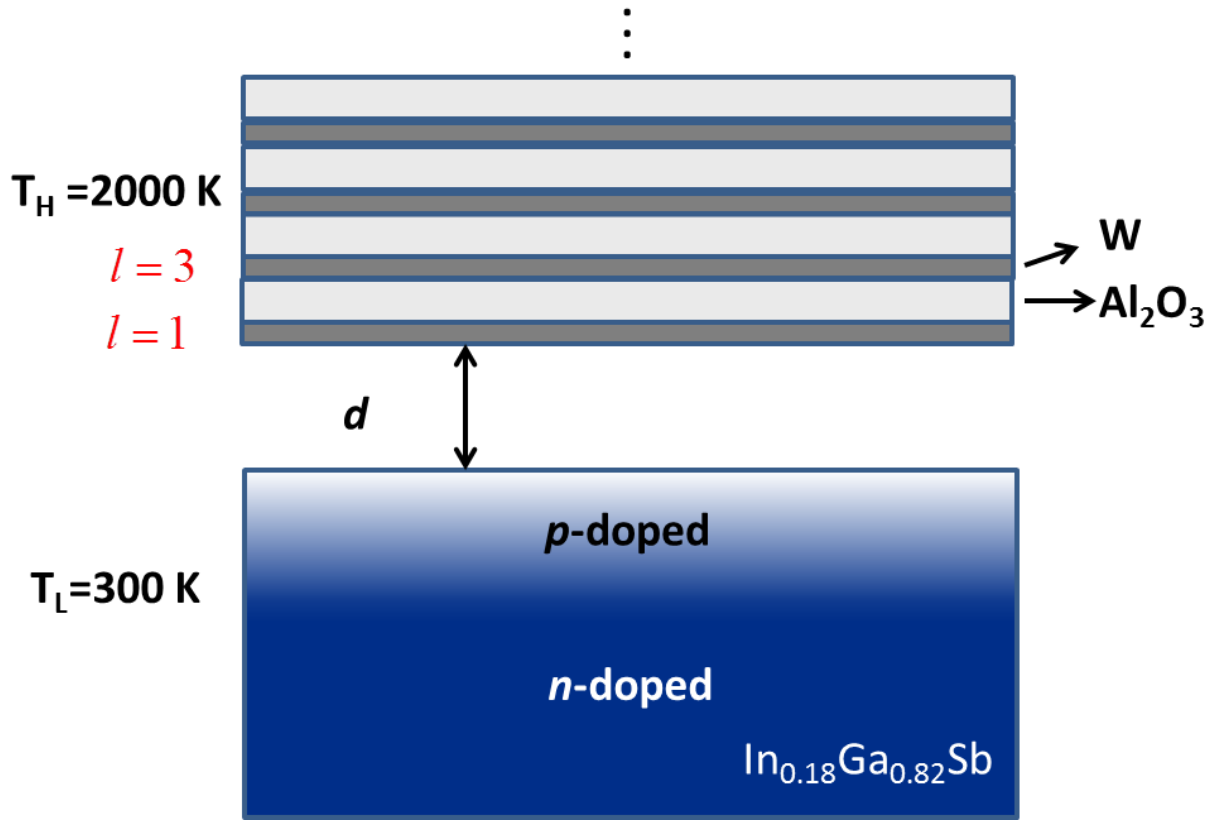

Fig. 1. The configuration of near-field TPV system when applying the multilayer emitter with alternate tungsten and alumina layer, and tungsten is the topmost layer. The TPV cell is made of the alloy $In_{0.18}Ga_{0.82}Sb$. The emitter and cell temperatures are set as 2000 K and 300 K, respectively.

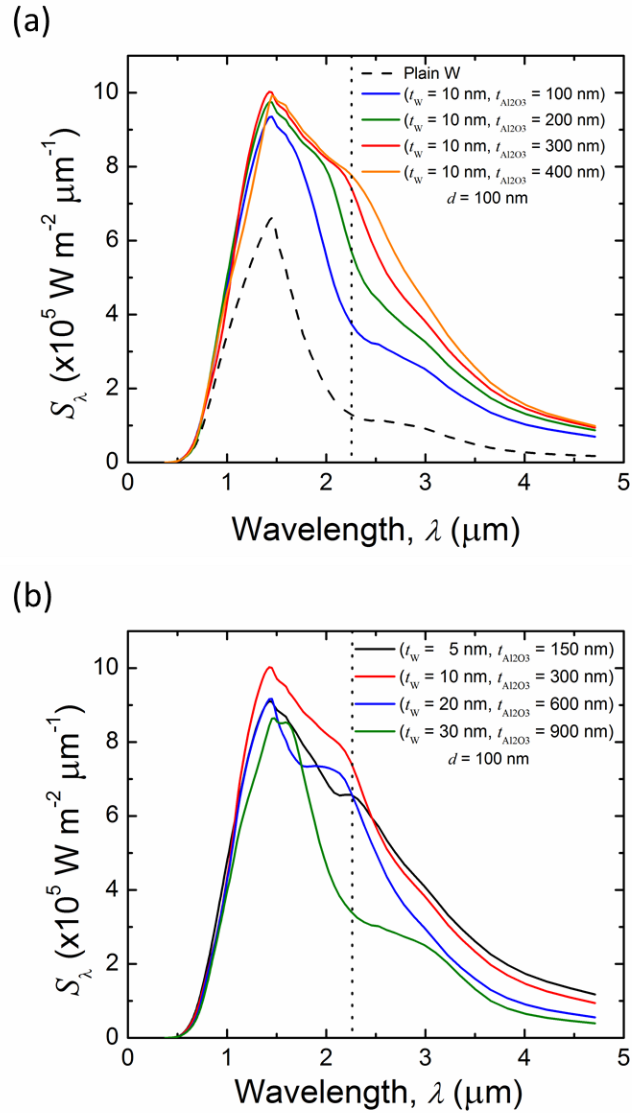

Fig. 2. Spectral heat fluxes between the multilayer emitter and the receiver (a) when the tungsten layer thickness is set as 10 nm, while the alumina layer thickness is varied; (b) when the ratio of tungsten layer thickness and alumina layer thickness is fixed as 1/30, while the tungsten layer thickness is different. The dash vertical line indicates the bandgap of TPV cell. The vacuum gap distance $d$ is 100 nm.

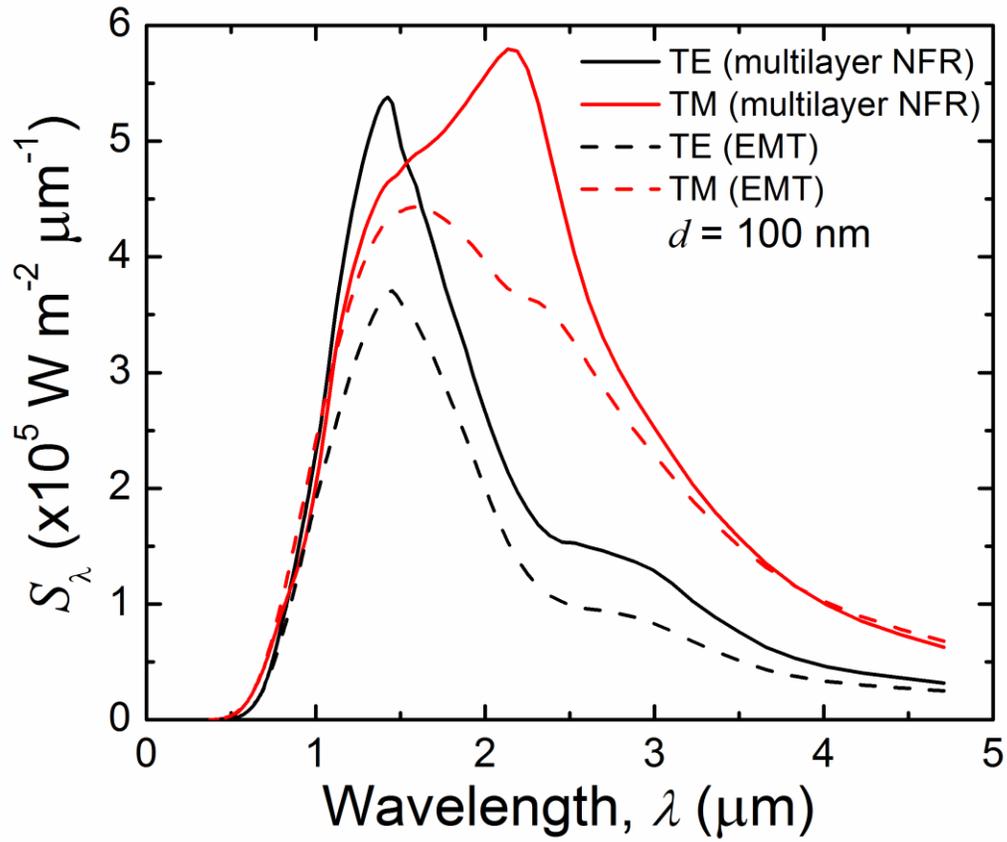

Fig. 3. Comparison of the spectral heat fluxes between the TPV cell and multilayer emitter using exact multilayer NFR calculation and EMT calculation in different polarizations of TE waves and TM waves. The vacuum gap distance is 100 nm.

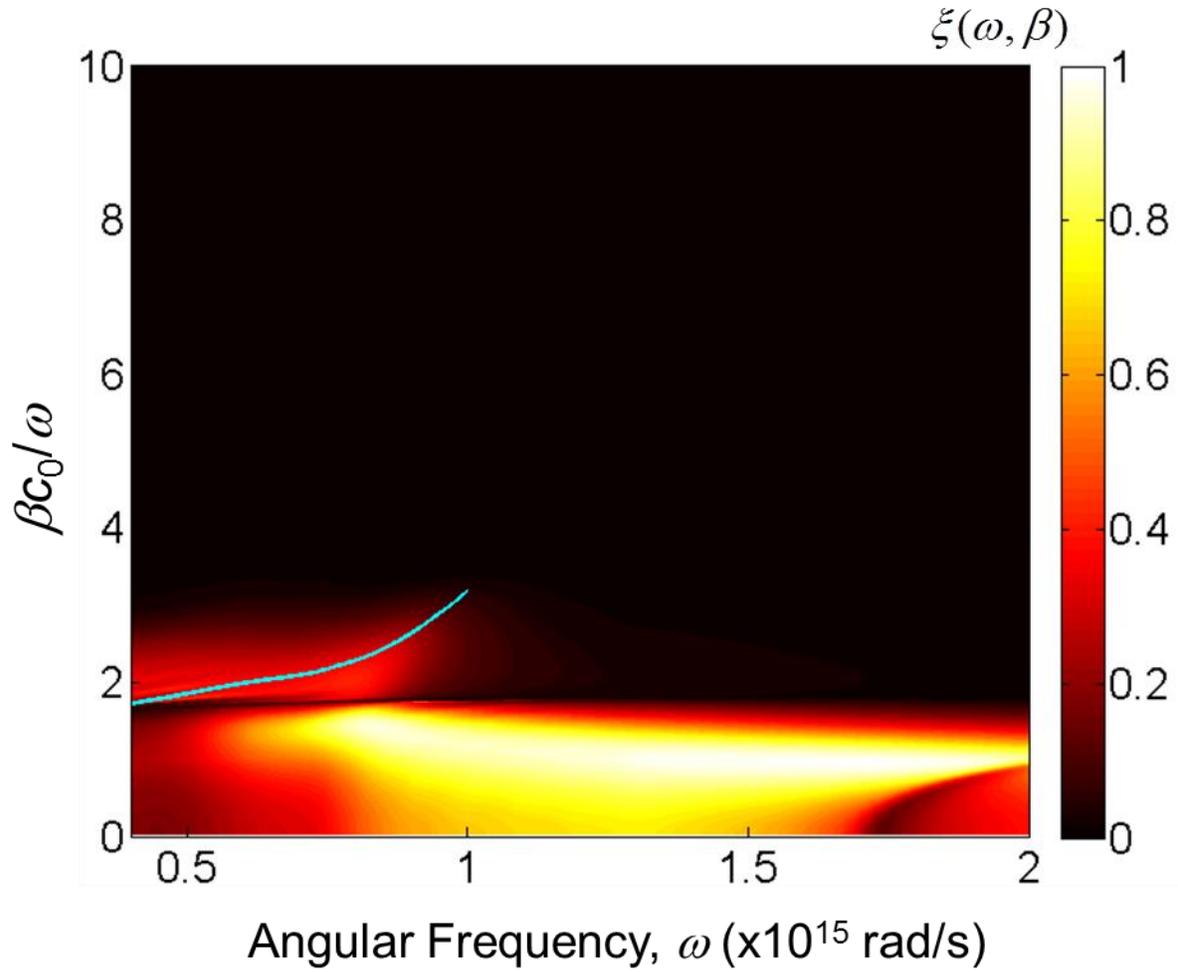

Fig. 4. The contour plots of energy transmission coefficient for TM waves between multilayer emitter and TPV cell using exact multilayer NFR calculation. The dispersion curve within tungsten thin film with substrates of vacuum and alumina on each side is also plotted. Note that the parallel wavevector is normalized to the wavevector in vacuum. The vacuum gap distance is 100 nm.

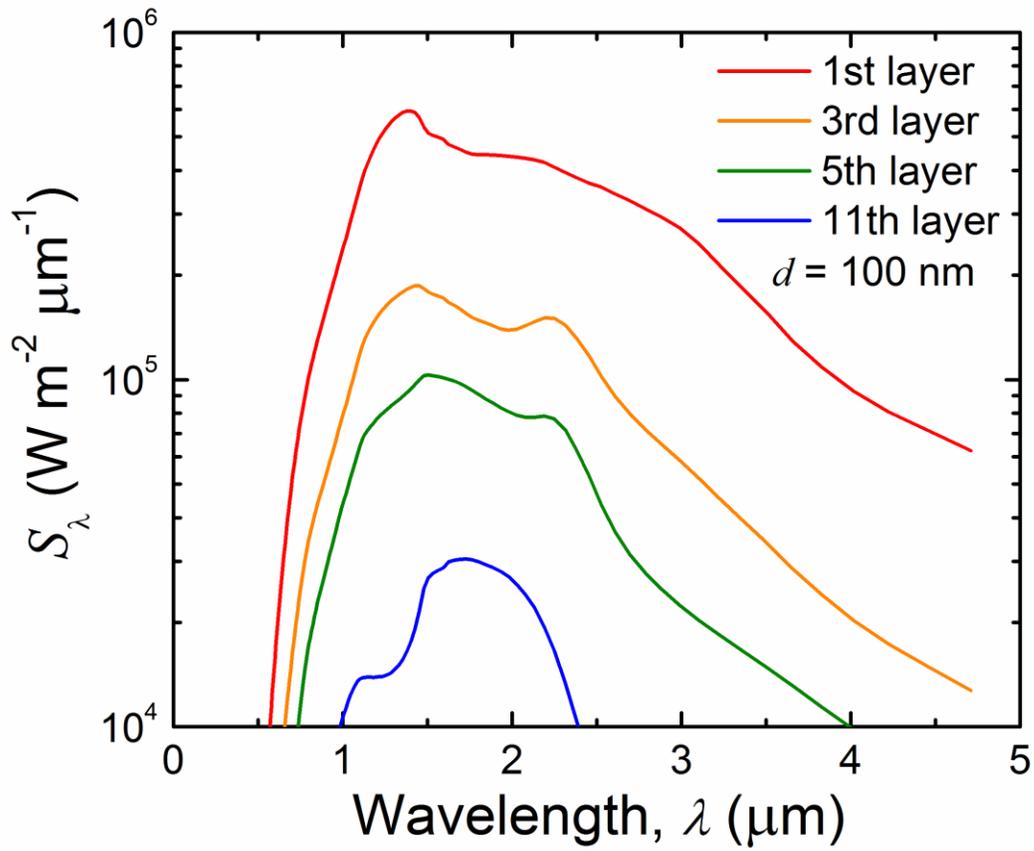

Fig. 5. The spectral heat flux emitted from different tungsten layer of the multilayer emitter. When considering the emission from a single layer, the other layers just function as the filters. The vacuum gap distance is 100 nm.

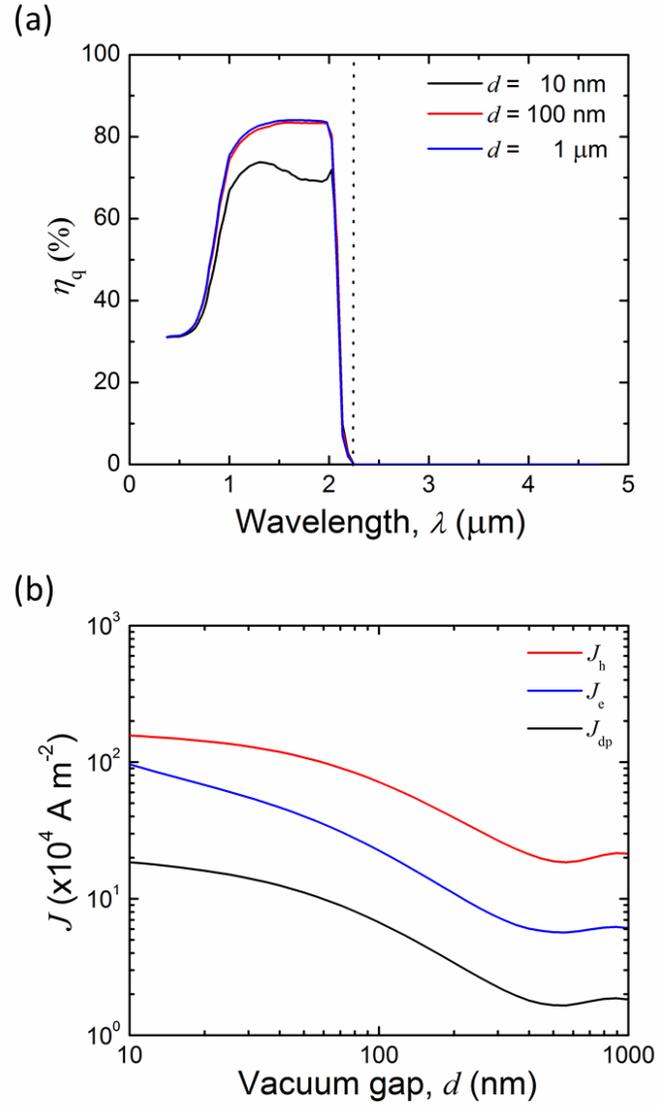

Fig. 6. (a) The quantum efficiency (QE) vs wavelength for different vacuum gap distances. (b) The photocurrent density generated in different regions of TPV cell against different vacuum gap distances.

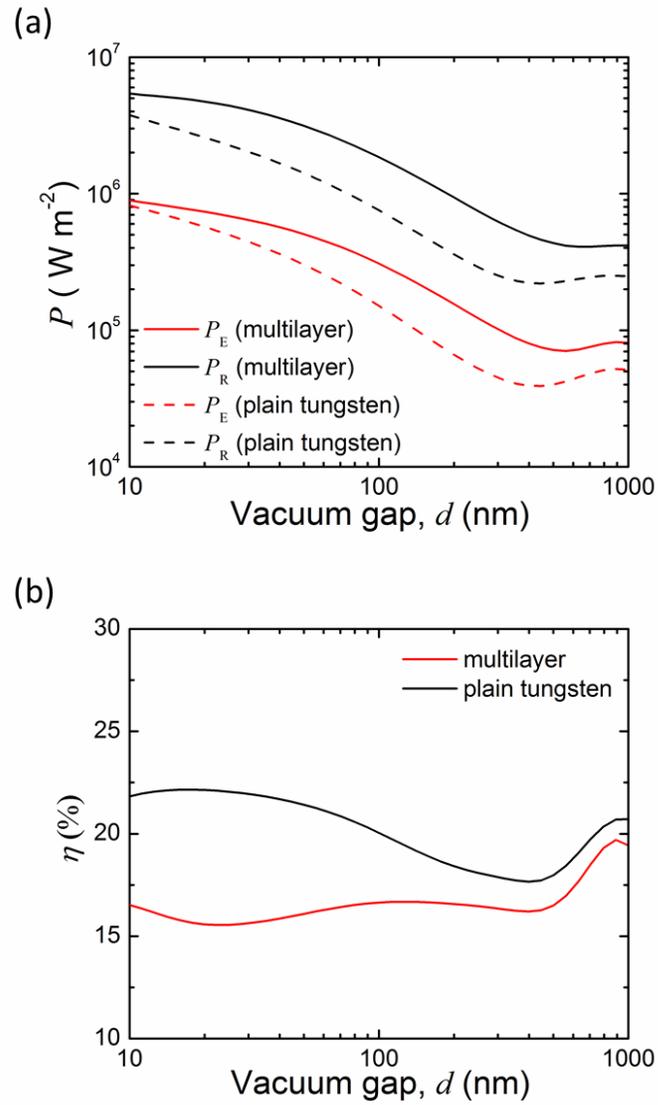

Fig. 7. (a) The electrical power output and radiative power input, and (b) the conversion efficiency of the TPV system for both multilayer emitter and plain tungsten emitter vs vacuum gap distance.

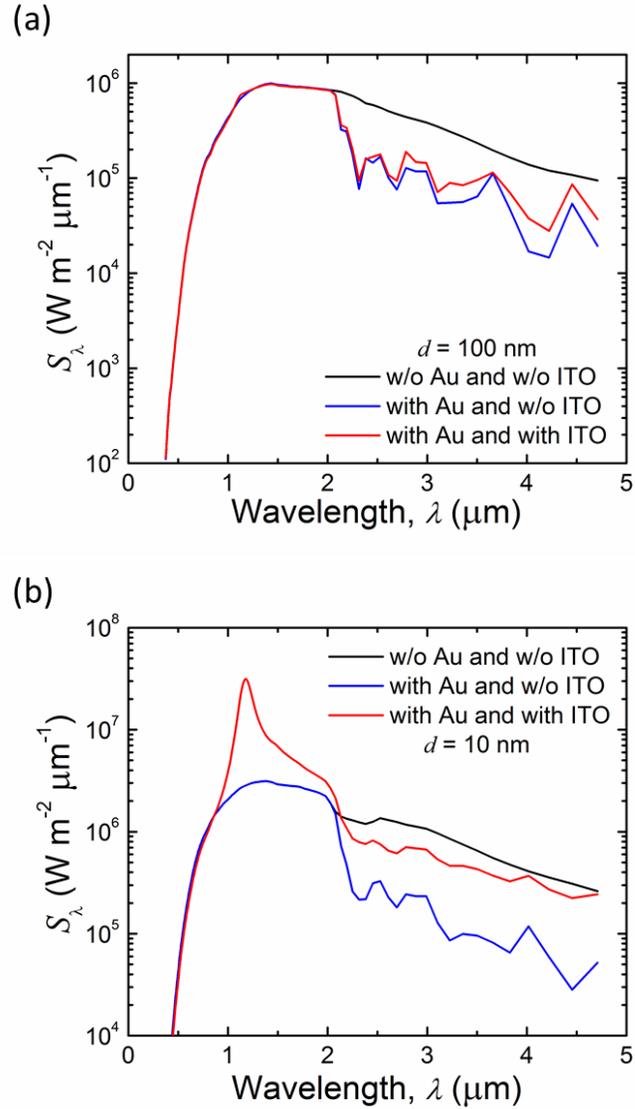

Fig. 8. (a) The spectral heat fluxes absorbed by the receiver vs wavelength at $d = 100$ nm and (b) $d = 10$ nm for three different cases, which are just semiinfinite TPV cell, TPV cell with opaque Au on the back, and TPV cell with both opaque Au on the back and 5-nm-thick ITO at the front, respectively.

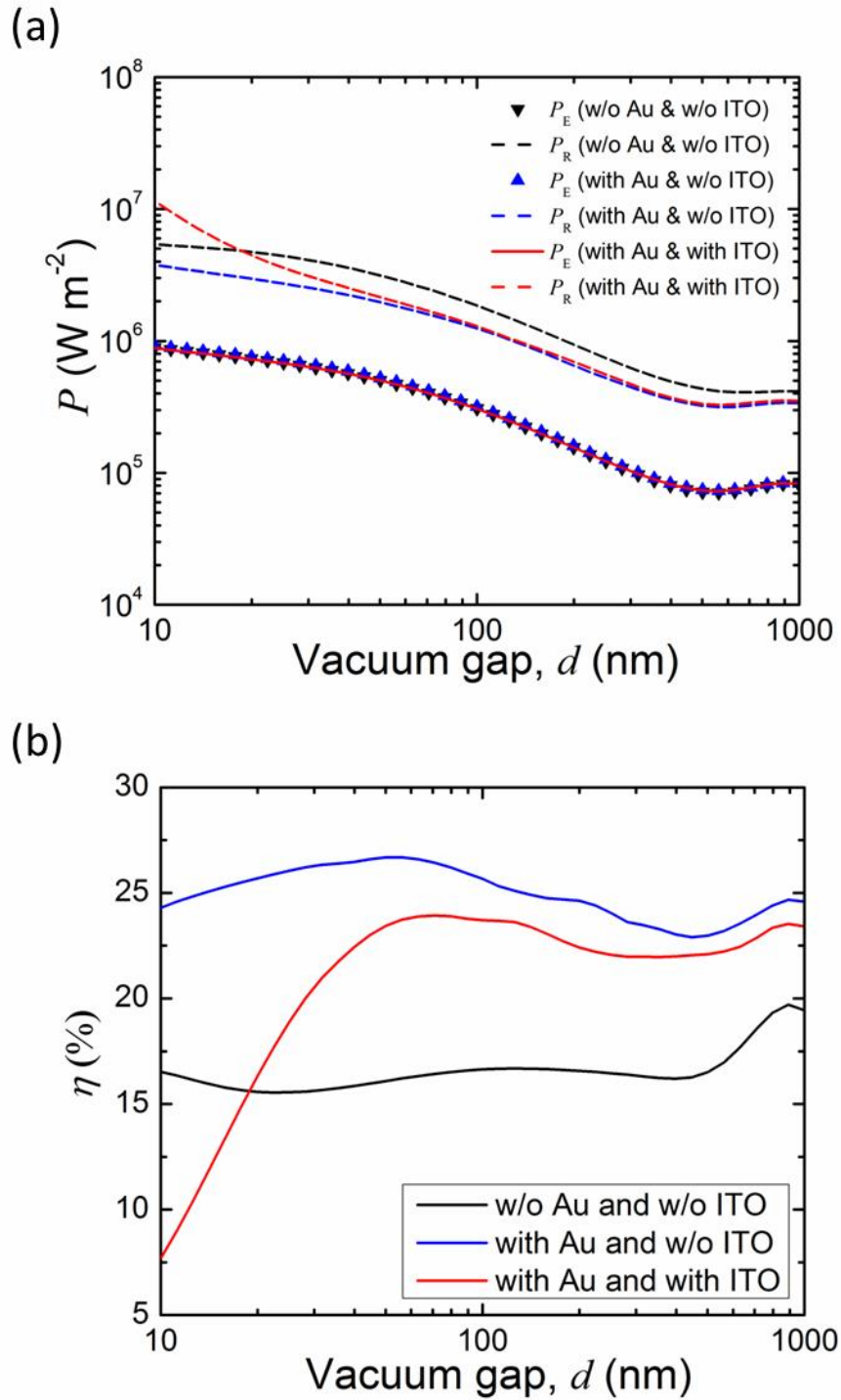

Fig. 9. (a) The radiative power input and electrical power output vs vacuum gap distance, and (b) the conversion efficiency vs vacuum gap distance for three different cases, which are just semiinfinite TPV cell, TPV cell with opaque Au on the back, and TPV cell with both opaque Au on the back and 5-nm-thick ITO at the front, respectively.